\newif\ifmnras

\ifmnras
\documentclass[a4paper,fleqn,usenatbib]{mnras}
\usepackage{newtxtext,newtxmath}
\usepackage[T1]{fontenc}
\usepackage{ae,aecompl}

\usepackage{graphicx}	
\usepackage[section] {placeins}
\usepackage{subfigure}
\usepackage{float}
\usepackage{color}
\usepackage{hyperref}
\graphicspath{{./figures/}}
\usepackage{float}

\def\msun{{\rm\,M_\odot}} 
\def\lsun{{\rm\,L_\odot}}
\def\zsun{{\rm\,Z_\odot}}
\newcommand{\etal}{et al.\ }
\newcommand{\kms}{\, {\rm km\, s}^{-1}}
\newcommand{\ikms}{(\kms)^{-1}}
\newcommand{\mpc}{\, {\rm Mpc}}
\newcommand{\kpc}{\, {\rm kpc}}
\newcommand{\hmpc}{\, h^{-1} \mpc}
\newcommand{\ihmpc}{(\hmpc)^{-1}}
\newcommand{\hkpc}{\, h^{-1} \kpc}
\newcommand{\lya}{Ly$\alpha$}
\newcommand{\lyaf}{Ly$\alpha$ forest}
\newcommand{\ch}{\bf change}
\newcommand{\gmo}{{\gamma-1}}
\newcommand{\bF}{\bar{F}}
\newcommand{\hi}{\mbox{H\,{\scriptsize I}\ }}
\newcommand{\heii}{\mbox{He\,{\scriptsize II}\ }}
\newcommand{\civ}{\mbox{C\,{\scriptsize IV}\ }}
\newcommand{\kpa}{k_\parallel}
\newcommand{\vk}{{\mathbf k}}
\newcommand{\df}{\delta_F}
\newcommand{\sF}{{F_s}}
\newcommand{\sdelta}{{\delta_s}}
\newcommand{\seta}{{\eta_s}}
\newcommand{\dt}{\Delta \theta}
\newcommand{\dv}{\Delta v}
\newcommand{\pa}{\parallel}
\newcommand{\pe}{\perp}
\newcommand{\dz}{\Delta z}
\newcommand{\llya}{L$_{{\rm Ly}\alpha}$}
\newcommand{\lheii}{L$_{{\rm He II}}$}
\newcommand{\lciv}{L$_{{\rm C IV}}$}
\newcommand{\expZ}{$\langle Z \rangle$}
\newcommand{\expT}{$\langle T \rangle$}
\newcommand{\expD}{$\langle n_{{\rm H}} \rangle$}
\newcommand{\expF}{$\langle f_{{\rm HI}} \rangle$}
\def\h2{${\rm\,H_2}$}

\title{Testing Models of Quasar Hosts With Strong Gravitational Lensing by Quasar Hosts}

\author[]{Renyue Cen$^{1}$ and Mohammadtaher Safarzadeh$^{2}$
\\
$^{1}$Department of Astrophysical Sciences, Princeton University, Princeton, NJ 08544\\
$^{2}$Department of Physics and Astronomy, Johns Hopkins University, Baltimore, MD 21218, USA 
} 

\pubyear{2015}

\begin{document}
\label{firstpage}
\pagerange{\pageref{firstpage}--\pageref{lastpage}}
\maketitle

\else

\pdfoutput=1
\documentclass[12pt,preprint]{aastex6}

\textheight=9.2in
\topmargin=-0.5in
\textwidth=6.5in
\rightmargin=2.0in

\usepackage[T1]{fontenc}
\usepackage{ae,aecompl}

\usepackage{graphicx}	
\usepackage[section] {placeins}
\usepackage{subfigure}
\usepackage{float}
\usepackage{color}
\usepackage{hyperref}
\graphicspath{{./figures/}}
\usepackage{float}

\usepackage[scaled]{helvet}
\renewcommand*\familydefault{\sfdefault}
\usepackage[T1]{fontenc}

\def\msun{{\rm\,M_\odot}} 
\def\lsun{{\rm\,L_\odot}}
\def\zsun{{\rm\,Z_\odot}}
\newcommand{\etal}{et al.\ }
\newcommand{\kms}{\, {\rm km\, s}^{-1}}
\newcommand{\ikms}{(\kms)^{-1}}
\newcommand{\mpc}{\, {\rm Mpc}}
\newcommand{\kpc}{\, {\rm kpc}}
\newcommand{\hmpc}{\, h^{-1} \mpc}
\newcommand{\ihmpc}{(\hmpc)^{-1}}
\newcommand{\hkpc}{\, h^{-1} \kpc}
\newcommand{\lya}{Ly$\alpha$}
\newcommand{\lyaf}{Ly$\alpha$ forest}
\newcommand{\ch}{\bf change}
\newcommand{\gmo}{{\gamma-1}}
\newcommand{\bF}{\bar{F}}
\newcommand{\hi}{\mbox{H\,{\scriptsize I}\ }}
\newcommand{\heii}{\mbox{He\,{\scriptsize II}\ }}
\newcommand{\civ}{\mbox{C\,{\scriptsize IV}\ }}
\newcommand{\kpa}{k_\parallel}
\newcommand{\vk}{{\mathbf k}}
\newcommand{\df}{\delta_F}
\newcommand{\sF}{{F_s}}
\newcommand{\sdelta}{{\delta_s}}
\newcommand{\seta}{{\eta_s}}
\newcommand{\dt}{\Delta \theta}
\newcommand{\dv}{\Delta v}
\newcommand{\pa}{\parallel}
\newcommand{\pe}{\perp}
\newcommand{\dz}{\Delta z}
\newcommand{\llya}{L$_{{\rm Ly}\alpha}$}
\newcommand{\lheii}{L$_{{\rm He II}}$}
\newcommand{\lciv}{L$_{{\rm C IV}}$}
\newcommand{\expZ}{$\langle Z \rangle$}
\newcommand{\expT}{$\langle T \rangle$}
\newcommand{\expD}{$\langle n_{{\rm H}} \rangle$}
\newcommand{\expF}{$\langle f_{{\rm HI}} \rangle$}
\def\h2{${\rm\,H_2}$}

\begin{document}
\label{firstpage}

\title{Testing Models of Quasar Hosts With Strong Gravitational Lensing by Quasar Hosts}

\author{Renyue Cen\altaffilmark{1} and Mohammadtaher Safarzadeh\altaffilmark{2}}

\footnotetext[1]{Department of Astrophysical Sciences, Princeton University, Princeton, NJ 08544; cen@astro.princeton.edu}

\footnotetext[2]{Johns Hopkins University, Department of Physics and Astronomy, Baltimore, MD 21218, USA}

\fi

\begin{abstract}

We perform a statistical analysis of strong gravitational lensing by 
quasar hosts of background galaxies, 
in the two competing models of dark matter halos of quasars, HOD and CS models.
Utilizing the {\em BolshoiP Simulation}
we demonstrate that strong gravitational lensing provides a potentially very powerful
test of models of quasar hosting halos.
For quasars at $z=0.5$, 
the lensing probability by quasars of background galaxies 
in the HOD model is higher than that of the CS model by 
two orders of magnitude or more for lensing image separations in the range of $\theta\sim 1.2-12~$arcsec.
To observationally test this, 
we show that, as an example, at the depth of the CANDELS wide field survey 
and with a quasar sample of $1000$ at $z=0.5$,
the two models can be differentiated at $3-4\sigma$ confidence level.

\end{abstract}




\section{Introduction}

The basic characteristics of the dark matter halos hosting quasars, such as their masses, remain uncertain.
The conventional, popular HOD (halo occupation distribution) model
stands out on its simplicity in that it is based on assigning a probability function to quasars
to reside in a halo of a given mass in order to match the observed quasar clustering strength \citep{2005Zheng,2007Zheng,2013Shen}.
A newly proposed model \citep[]['CS model' hereafter]{2015Cen} differs in its physical proposition
by allowing for considerations of physical conditions of gas in galaxies hosting quasars 
and the fact that quasar activities are a special, rare phase.
In particular, a restriction (an upper bound) on the halo mass of quasar hosts of $10^{12.5-13}\msun$ is imposed as a necessary condition,
based on the physical condition that more massive halos, being completely hot gas dominated, 
are incapable of feeding the central supermassive black holes in a vigorous fashion.
In addition, since the typical quasar duty cycle is much less than unity, 
quasar activity must not be a typical condition in a galaxy's history.
We thus argue that some special condition is necessary to ``trigger" each quasar feeding event.
We propose that significant gravitational interaction, 
parameterized as the presence of a significant companion halo within some distance,
not necessarily a major merger,
is the second necessary condition for making a quasar.
We demonstrate that the CS model can equally well match the observed quasar clustering
properties (auto-correlation functions of quasars and quasar-galaxy cross-correlation functions)
over a wide range of redshift.

The masses of the dark matter halos in the CS model are very different from those of the HOD based model.
For example, at $z\sim 0.5-2$, the host halos in the CS model have masses of $\sim 10^{11}-10^{12}\msun$,
compared to $\ge 10^{13}\msun$ in the HOD model.
This large difference gives rise to important differences in several observables.
First, a critical differentiator is the cold gas content in quasars host galaxies,
which is in part, but not entirely, due to the physical ingredients used in the construction of our model, 
i.e., the upper halo mass limit.
Specifically, because of the large halos mass required in the HOD model,
quasars hosts have much lower content of cold gas than in the CS model.
\citet[][]{2015Cen} have shown that
the CS model is in excellent agreement with the observed 
covering fraction of $60\%-70\%$ for Lyman limit systems within the virial radius of $z\sim 2$ quasars \citep[][]{2013Prochaska}.
On the other hand, the HOD model is
inconsistent with observations of the high covering fraction of Lyman limit systems in quasar host galaxies. 
Second, in \citet[][]{2015bCen} we show that,
while both HOD and CS models are consistent with 
the observed thermal Sunyaev-Zeldovich (tSZ) effect 
at the resolution of FWHM=$10~$arcmin obtained by Planck data,
FWHM=$1~$arcmin beam tSZ measurements would provide
a potentially powerful test between the two models.
Subsequently, a careful analysis of the South Pole Telescope tSZ data at a beam of FWHM$\sim 1~$arcmin
suggests that the CS model is strongly favored over the HOD model \citep[][]{2016Spacek}.
In this {\it Letter} we present and demonstrate yet 
another potentially powerful test to distinguish between these two competing models,
namely strong gravitational lensing by quasar hosting galaxies of background galaxies,
which may finally put to rest the issue of the halo masses of quasar hosts.

\section{Simulations and Analysis Method}\label{sec: sims}

We utilize the {\em Bolshoi Simulation} \citep[][]{2011Klypin} to perform the analysis.
A set of properties of this simulation that meet our requirements includes
a large box of $250h^{-1}$Mpc,
a relatively good mass resolution with dark matter particles of mass $1.3 \times 10^8 h^{-1}\msun$,
and a spatial resolution of 1 $h^{-1}\,{\rm kpc}$ comoving.
The mass and spatial resolutions are adequate for capturing halos of masses greater than $2\times10^{10}\msun$,
which are resolved by at least about 100 particles and 40 spatial resolution elements for the virial diameter.
Since the mass range of interest here is $\ge 10^{11}\msun$,
all halos concerned are well resolved.
Dark matter halos are found through a {\em friends-of-friends} (FOF) algorithm.
The adopted $\Lambda{\rm CDM}$ cosmology parameters are
$\Omega_{\rm m}=0.27$, $\Omega_{\rm b}=0.045$, $\Omega_{\Lambda}=0.75$,
$\sigma_8=0.82$ and $n=0.95$, where the Hubble constant is $H_0=100{h\,\rm km}
\,{\rm s}^{-1}\,{\rm Mpc}^{-1}$ with $h=0.70$.

We select quasars host halos from $z=0.5$ data output of the Bolshoi Simulation,
using the detailed prescriptions for both CS and HOD models, described in \citet[][]{2015Cen}. 
For the purpose of computing lensing statistics, 
we project all particles in the $z=0.5$ simulation box along the x-axis onto a plane with spatial resolution of $4$ proper kpc.
At the location of each quasar halo, we compute the 
radial profile of the projected dark matter density centered on the halo. 
In addition to the dark matter, we also model the baryons' contribution to
the projected surface density. Following the parametrization 
of \citet{2013Behroozi}, we assign a baryon fraction to the dark matter halos as a function of the halo mass. 
The baryon mass is distributed and projected assuming a Singular Isothermal Sphere (SIS) model. 
The SIS radius for the baryons is defined to be ${\rm r_{SIS}=2\times r_{eff}}$ where the effective radius is computed 
following \citet{2014vanderWel} fits for elliptical galaxies in that redshift range:
\begin{equation}
\label{eq:graph_eq}
{\rm r_{eff}= 10^{0.78}(\frac{M}{5\times 10^{10}\msun})^{0.22}}
\end{equation}
\noindent
The projected surface density, without and with baryonic correction,
subtracted by the mean surface density of the box (${\rm \sim 3\times 10^7 M_{\odot}/kpc^2}$),
is compared to critical surface density. 
We compute the surface density of the halos in radial bins 
and the radius within which the mean surface density is equal to $\Sigma_{crit}$  
is defined as the Einstein radius $r_E$ for that halo,
with the corresponding angle subtended being $\theta_E=r_E/D_L$,
where $D_L$ is the angular diameter distance to the quasar host (i.e., the lens).
The critical density for strong lensing is
\begin{equation}
\label{eq:graph_eq}
\Sigma_{crit}= \frac{c^2}{4 \pi G}\frac{D_S}{D_L D_{LS}},
\end{equation}
\noindent
at redshift $z_l$; in this paper, we consider $z_L=0.5$ for illustration.
Here $D_S$ is the angular diameter
distance to the source and $D_{LS}$ is the angular diameter distance between the lens and the source.
We note that for sources at $z_{s}>2$ the critical density for strong lensing 
by a lens at $z_l=0.5$ is approximately constant ${\rm \sim 10^{9} M_{\odot}/kpc^2}$,
which rises slowly to ${\rm \sim 2\times10^{9} M_{\odot}/kpc^2}$, at $z_s=1$, followed by a steep rise to $z_s=0.5$.
We compute the surface densities of 10,000 quasar host halos 
for both CS and HOD models and obtain 
the probability distribution function (PDF) of the lensed image angular separation statistics for each model.

In order to make quantitative calculations for lensing statistics of background galaxies,
we assess the lensing cross section in the source plane as follows.
We assume an SIS model for the lens in which all the sources in the background
whose {\em un-deflected} photons pass within the lens's $r_E$ are lensed to give two images,
with the cross section in the source plane giving two images being 
$\sigma=\pi r_{E}^2$ \citep{1984Turner}.
Defining the impact parameter as ($f\equiv\theta_Q/\theta_E$),
lensing of background galaxies with $f<1$ gives two images.
The amplification as a function of impact parameter is $r=\frac{1+f}{1-f}$.
Averaging the amplification over the cross section gives a factor of four
total amplification due to lensing inside the critical radius. 
In our case, we demand that both images are observed in order for us to be sure of a strong lensing event.
With that requirement, we find that, in a magnitude limited survey,
for sources within a given redshift interval,
the effective source plane galaxy number density turns out to be unchanged.
In other words, although
we can probe to fainter limits because of the total amplification power, 
the number of pairs of images both detectable is unchanged,
with the effective source plane across section remaining at $\sigma=\pi r_{E}^2$.

Then, we obtain the number of multiply imaged galaxies as
function of 
image separation $\Delta\theta=2\theta_E$ for a given $\Sigma_{gal}$ for each model.
We compare the distributions of $\Delta\theta$ between the HOD and CS models
We compute $\chi^2$ to statistically evaluate the size of quasar samples and the number density of background galaxies
required in order to differentiate between the HOD and CS models, using Poisson statistics.

The difference between the models for each radial bin is computed as follow:
\begin{equation}
\label{eq:sigmai}
\sigma_{i}^2= \frac{(N_{CS,i} - N_{HOD,i} )^2}{N_{CS,i} + N_{HOD,i} }
\end{equation}
for $i$ denotes the radial bin and 
$N_{CS,i} = \Sigma_{gal}\times2\pi r_i dr \times N_{QSO}\times P_{CS,i}$, where $P_{CS,i}$ 
is the probability of the CS model in {\em ith} radial bin at $r_i$. The same is adopted for HOD model. 
The total difference taking into account all the radial bins is computed as follow and shown in 
Figure~\ref{fig:chi2} below.

\begin{equation}
\label{eq:sigmat}
\sigma_{tot}^2= \sum \sigma_{i}^2
\end{equation}

\section{Results}

\ifmnras

\begin{figure}
\centering
\vskip 0.0cm
\includegraphics[width=3.5in]{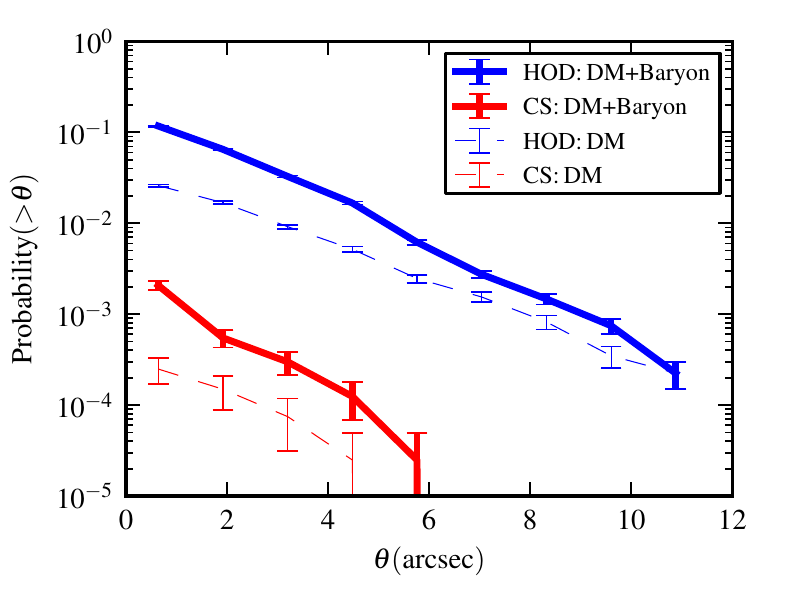}
\vskip -0.2cm
\caption{
shows the cumulative probability distribution functions of image separations $\theta$
in the HOD (blue curves) and CS (red curves) models,
without (dashed curves) and with (solid curves) baryons, respectively.
The result is based on 13,000 quasar host halo candidates 
in each model at z=0.5, viewed along each of the three orthogonal directions,
resulting in a total of 39,000 effective candidates.    
The errorbars are based on Poisson statistics.
}
\label{fig:hist}
\end{figure}

\else

\begin{figure}[!h]
\centering
\vskip -0.0cm
\resizebox{4.5in}{!}{\includegraphics[angle=0]{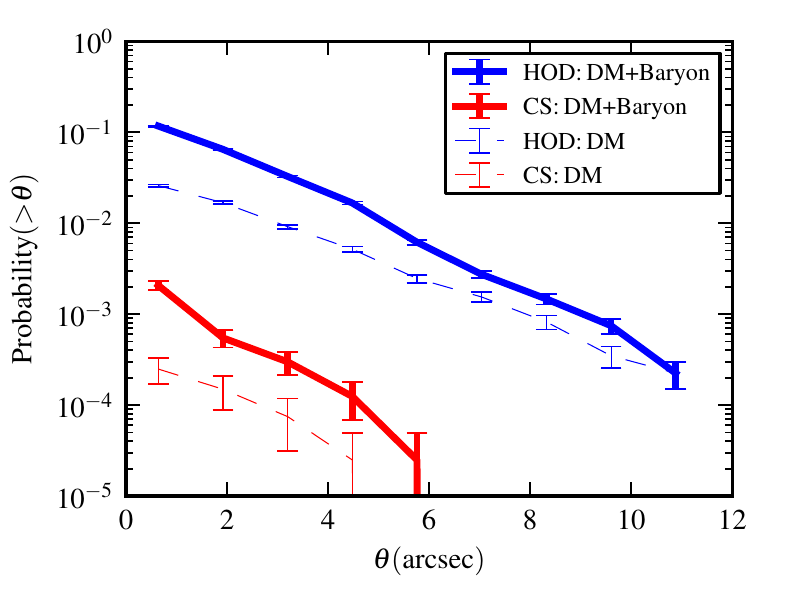}}
\vskip -0.0cm
\caption{
shows the cumulative probability distribution functions of image separations $\theta$
in the HOD (blue curves) and CS (red curves) models,
without (dashed curves) and with (solid curves) baryons, respectively.
The result is based on 13,000 quasar host halo candidates 
in each model at z=0.5, viewed along each of the three orthogonal directions,
resulting in a total of 39,000 effective candidates.    
The errorbars are based on Poisson statistics.
}
\label{fig:hist}
\end{figure}

\fi

Figure~\ref{fig:hist} 
shows the cumulative probability distribution function of image separations
in the HOD (blue curves) and CS (red curves) models,
without (dashed curves) and with (solid curves) baryons,
respectively.
We see that the large difference in masses of quasar host halos between HOD and CS models
is most vividly displayed:
the lensing probability in the HOD model is higher than that of the CS model by 
two orders of magnitude or more over the range $\theta\sim 1-5$ arcsec.
Above $\sim 6$ arcsec image separation there is no case in the CS model,
whereas the lensing probability in the HOD model is still at $\sim 10^{-4}-10^{-3}$ at $\sim 10-12$ arcsec.
We note that $1$ arcsec corresponds to $6.2$kpc at $z=0.5$.
The pixel size of the mass projection map at $z=0.5$ 
corresponds to $0.65$ arcsec in angular size.  
Thus, we do not include bins at $\theta<1.2$ arcsec in our considerations of differentiations
between the two models.

\ifmnras

\begin{figure}
\centering
\vskip 0.0cm
\includegraphics[width=3.5in]{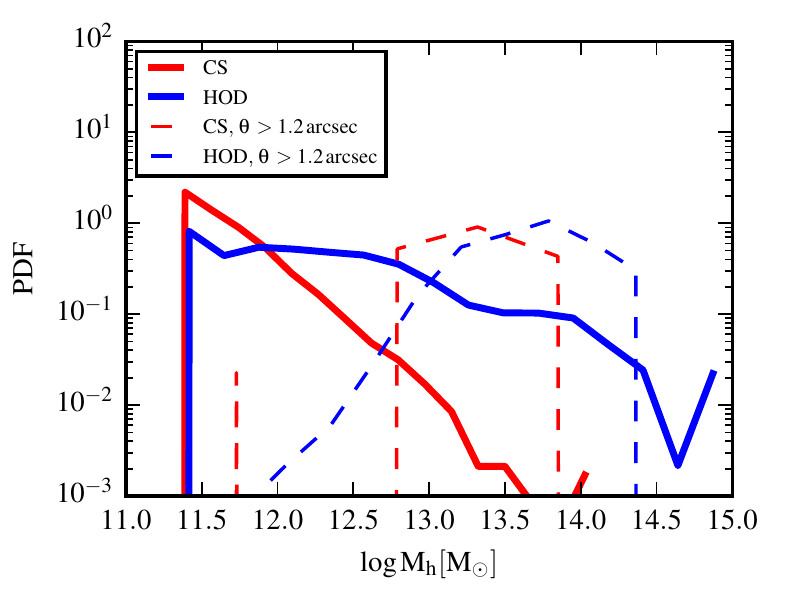}
\vskip -0.2cm
\caption{
shows the normalized probability distribution functions of quasar host halo masses
in the HOD (blue solid curve) and CS (red solid curve) models, respectively,
based on 13,000 halos used for Figure~\ref{fig:hist}.
The corresponding dashed curves 
are the normalized probability distribution functions of masses of selected quasar host halos
capable of producing strong lensing with image separation $\theta>1.2$~arcsec.
It is useful to note that the quasars at $z\sim 0.5$ that the models model 
have bolometric luminosity threshold of $10^{45.1}$erg/s \citep[][]{2015Cen}.
}
\label{fig:masshist}
\end{figure}

\else

\begin{figure}[!h]
\centering
\vskip -0.0cm
\resizebox{4.5in}{!}{\includegraphics[angle=0]{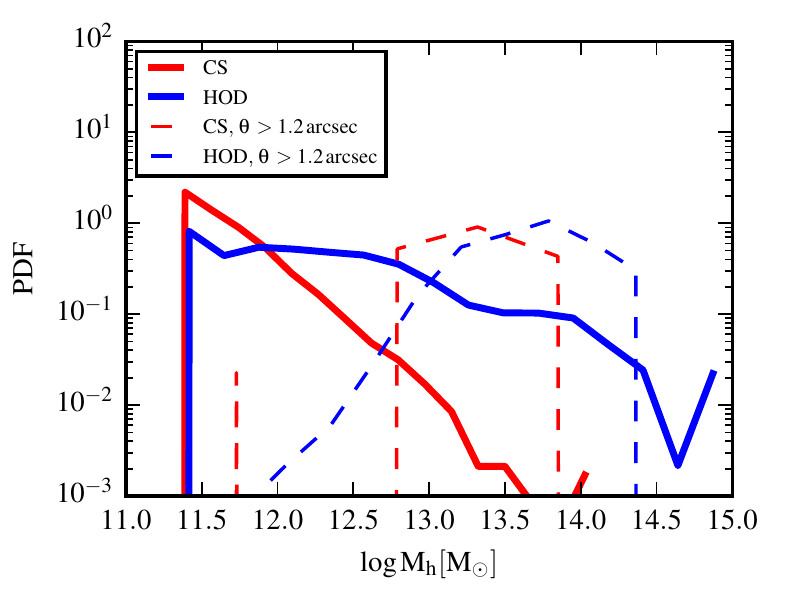}}
\vskip -0.0cm
\caption{
shows the normalized probability distribution functions of quasar host halo masses
in the HOD (blue solid curve) and CS (red solid curve) models, respectively,
based on 13,000 halos used for Figure~\ref{fig:hist}.
The corresponding dashed curves 
are the normalized probability distribution functions of masses of selected quasar host halos
capable of producing strong lensing with image separation $\theta>1.2$~arcsec.
It is useful to note that the quasars at $z\sim 0.5$ that the models model 
have bolometric luminosity threshold of $10^{45.1}$erg/s \citep[][]{2015Cen}.
}
\label{fig:masshist}
\end{figure}

\fi

The large differences between the HOD and CS models shown in Figure~\ref{fig:hist},
can be understood by looking at Figure~\ref{fig:masshist}, 
which shows a comparison between the normalized probability distribution functions of masses of all quasar host halos
(solid curves) and of those capable of producing strong lensing with image separation $\theta>1.2$~arcsec
(dashed curves).
We see that the vast majority of quasar host halos 
producing strong lensing with image separations $\theta>1.2$~arcsec
have masses greater than $10^{12.5}\msun$, peaking 
at $10^{13}-10^{13.5}\msun$.
Even though the overall number of quasar hosts are the same in the two models,
their abundances for halos of masses around the peak 
($10^{13}-10^{13.5}\msun$) differ by about two orders of magnitude,
which evidently can account for most of the differences between the two 
lensing probabilities seen in Figure~\ref{fig:hist}.
There may be other conceivable differences, such as the density slopes in the central regions,
possibly due for example to difference residing environments of halos of the same masses,
between the two quasar host halos in the two models.
But as a whole, these other possible differences, if any,
do not appear to make a large difference to the overall lensing probability.

\ifmnras

\begin{figure}
\centering
\includegraphics[width=3.5in, keepaspectratio]{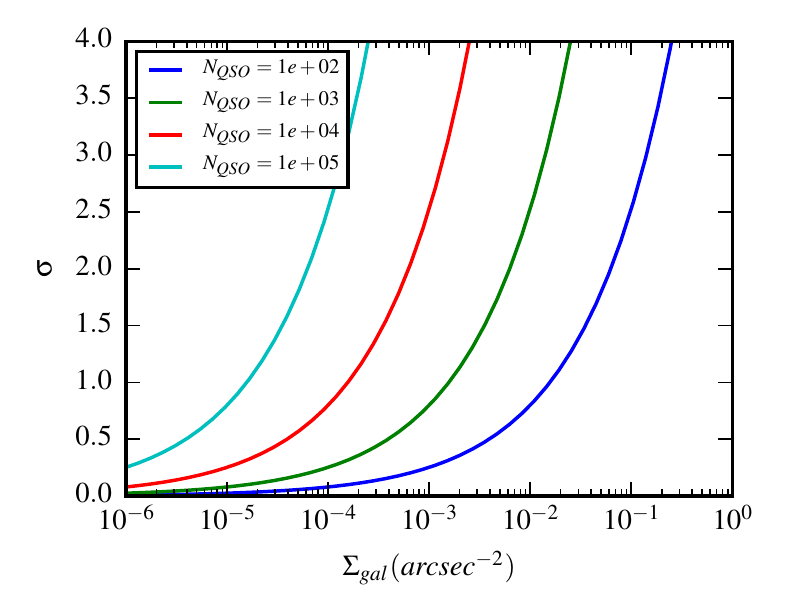}
\vskip 0.0cm
\caption{
Shows the confidence levels of statistical differentiation between the two models
as a function of the surface density of background galaxies, $\Sigma_{gal}$ (in arcsec$^{-2}$),
based on Eq~(\ref{eq:sigmai},\ref{eq:sigmat}).
Four cases of $N_{QSO}$ are shown.
We assume that the quasar hosts are at redshift $z=0.5$.
$N_{QSO}$ is the number of target lenses (i.e., quasar hosts).
Poisson statistics are used for the errorbars. 
}
\label{fig:chi2}
\end{figure}

\else

\begin{figure}[!h]
\centering
\vskip -0.0cm
\resizebox{4.5in}{!}{\includegraphics[angle=0]{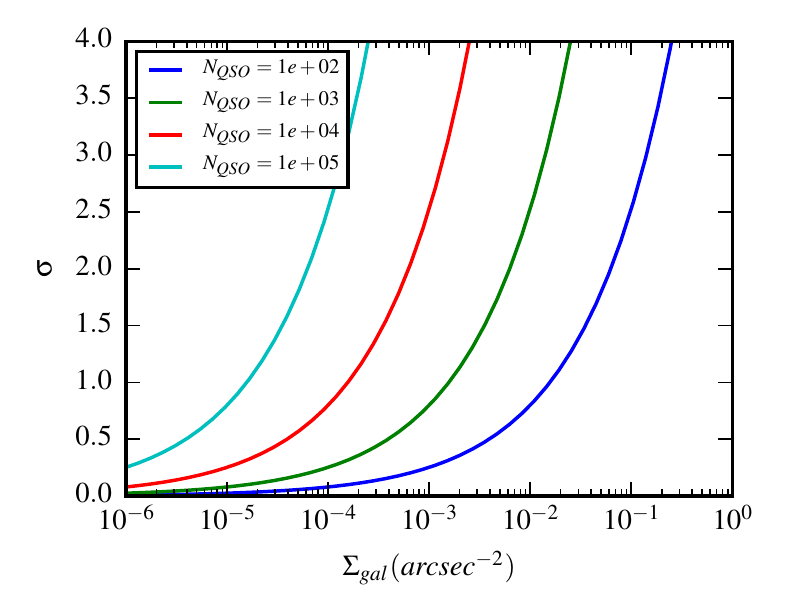}}
\vskip -0.0cm
\caption{
shows the confidence levels of statistical differentiation between the two models
as a function of the surface density of background galaxies, $\Sigma_{gal}$ (in arcsec$^{-2}$),
based on Eq~(\ref{eq:sigmai},\ref{eq:sigmat}).
Four cases of $N_{QSO}$ are shown.
We assume that the quasar hosts are at redshift $z=0.5$.
$N_{QSO}$ is the number of target lenses (i.e., quasar hosts).
Poisson statistics are used for the errorbars. 
}
\label{fig:chi2}
\end{figure}

\fi

\ifmnras

\begin{figure}
\centering
\includegraphics[width=3.5in, keepaspectratio]{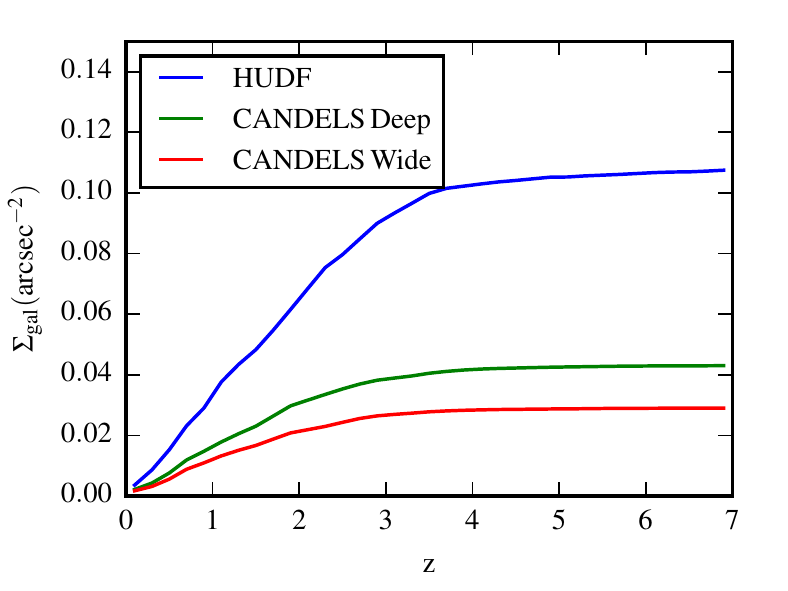}
\vskip 0.0cm
\caption{
Shows the cumulative surface number density of galaxies observed in the Hubble F160W filter 
down to 50\% completeness level in HUDF \citep{2006Beckwith} and CANDELS 
deep and shallow tier observations \citep{2011Grogin,2011Koekemoer}. 
The completeness level at 50\% corresponds to 
${\rm m_{AB}(F160W)}$ = 25.9, 26.6, 28.1 
for the CANDELS wide, deep and HUDF, respectively.
Data is from the compilation of \citet{2013Guo}. 
}
\label{fig:Sigma}
\end{figure}

\else

\begin{figure}[!h]
\centering
\vskip -0.0cm
\hskip -1.0cm
\resizebox{4.5in}{!}{\includegraphics[angle=0]{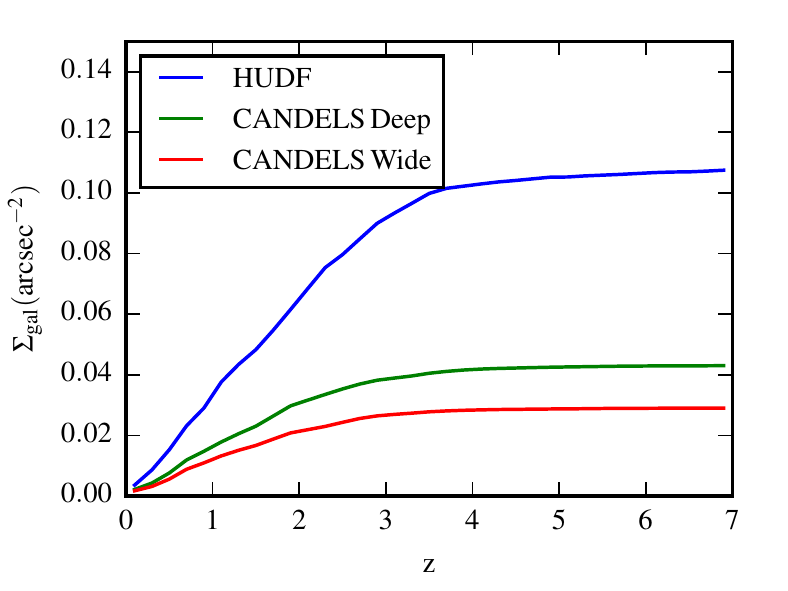}}
\vskip -0.0cm
\caption{
Shows the cumulative surface number density of galaxies observed in the Hubble F160W filter 
down to 50\% completeness level in HUDF \citep{2006Beckwith} and CANDELS 
deep and shallow tier observations \citep{2011Grogin,2011Koekemoer}. 
The completeness level at 50\% corresponds to 
${\rm m_{AB}(F160W)}$ = 25.9, 26.6, 28.1 
for the CANDELS wide, deep and HUDF, respectively.
Data is from the compilation of \citet{2013Guo}. 
}
\label{fig:Sigma}
\end{figure}

\fi

Given the results shown in 
Figure~\ref{fig:hist}, we now estimate the observational samples,
a combination of the number of target lenses (i.e., quasar hosts), $N_{QSO}$,
and the surface density of background galaxies, $\Sigma_{gal}$ (in arcsec$^{-2}$),
that are required to differentiate between the CS and HOD models.
To be specific, we assume that the quasar hosts are at redshift $z=0.5$.
Figure~\ref{fig:chi2} 
shows the confidence levels of statistical differentiation between the two models,
based on Eq~(\ref{eq:sigmai},\ref{eq:sigmat}).
We note that the results only depend on the product $N_{QSO}\Sigma_{gal}$
but we show four separate cases of $N_{QSO}$ for ease of assessment.
For example, for a quasar sample of $100$, 
a surface density of background galaxies of $\Sigma_{gal}= 0.1$ arcsec$^{-2}$
will allow for a $2.5\sigma$ test between the two models.
For a quasar sample of $1000$, 
$\Sigma_{gal}= 0.023$ arcsec$^{-2}$ produces a $4\sigma$ test.

To illustrate the observational feasibility of testing the models,
Figure~\ref{fig:Sigma} shows 
the cumulative surface number density of galaxies observed in the Hubble F160W filter 
down to 50\% completeness level in HUDF \citep{2006Beckwith} (blue curve) and CANDELS deep (green curve) and shallow (red curve)
tier observations \citep{2011Grogin,2011Koekemoer}. 
Numerically, we see that $\Sigma_{gal}(>z=1-2) \sim 0.01-0.02~$ arcsec$^{-2}$ 
for the CANDELS wide field survey; a survey of this depth with $1000$ quasar at $z=0.5$
would be able to differentiate between the two models at $\sim 3-4\sigma$ confidence level.
At the depth of HUDF $\Sigma_{gal}(>z=1-2) \sim 0.05-0.08$ arcsec$^{-2}$,
which could yield $\ge 2\sigma$ confidence level 
with only about $200$ quasars at $z=0.5$.

\section{Conclusions}

We perform a statistical analysis of strong gravitational lensing by  
quasar hosts of background galaxies, utilizing {\em BolshoiP Simulation}.
We demonstrate that strong gravitational lensing provides a potentially very powerful
test of models of quasar hosting halos.
Our focus is at $z=0.5$, where the difference in the halo masses of quasar hosts between competing models
is large and where the placement of lenses is near optimal for lensing of high redshift galaxies.

Our initial expectation that the large 
difference in masses of quasar host halos between HOD 
\citep{2005Zheng,2007Zheng,2013Shen}
and CS \citep[][]{2015Cen} model 
- 
a threshold mass of $(1-3)\times 10^{11}\msun$ in the CS model verus a median halo mass 
of $6\times 10^{12}\msun$ in the HOD model at $z\sim 0.5$ -
should be strongly discernible in strong lensing statistics is clearly borne out.
We find that the lensing probability in the HOD model is higher than that of the CS model by 
two orders of magnitude or more for lensing image separations in the range of $\theta\sim 1-5~$arcsec.
Above $\sim 6$ arcsec image separation there is no case in the CS model,
whereas the lensing probability in the HOD model is still at $\sim 10^{-4}-10^{-3}$ at image
separation of $\sim 10-12$ arcsec.

Translating this large theoretical difference between HOD and CS models into observables,
we show that, as an example,
at the depth of the CANDELS wide field survey 
and with a quasar sample of $1000$ at $z=0.5$,
the two models can be differentiated at $\sim 3-4\sigma$ confidence level.
The overall statistical power depends on the product $N_{QSO}\Sigma_{gal}$,
where $N_{QSO}$ is the quasar sample size and 
$\Sigma_{gal}$ is the surface density of detectable background galaxies.
In a pioneering observational study, 
\citet[][]{2012Courbin} report three cases of QSO lenses at $z\sim 0.2-0.3$
with velocity dispersion of the QSO hosts in the range of $210-285\kms$.
It is likely that, with a concerted effort, strong gravitational lensing by quasars may provide
the most definitive and direct test of host halo models for quasars.

\vskip 1cm
We are grateful to Anatoly Klypin for providing us with the projected mass maps
of Bolshoi-Planck simulation in a most prompt fashion. 
We thank Tommaso Treu for useful discussion.
This work is supported in part by grants NNX12AF91G and AST15-15389.










\label{lastpage}
\end{document}